\RequirePackage{fix-cm}
\documentclass[twocolumn,epjc3]{svjour3}  
\smartqed  

\usepackage[pdftex]{graphicx}

\usepackage{amssymb, amsmath}
\usepackage{bm}
\usepackage{ulem}
\RequirePackage[numbers,sort&compress]{natbib}
\newcommand{\lgle}{\left\langle}
\newcommand{\rgle}{\right\rangle}
\newcommand{\MAX}{\mbox{\scriptsize MAX}}
\newcommand{\eff}{\mbox{\scriptsize eff}}
\newcommand{\mini}{\mbox{\scriptsize min}}
\newcommand{\maxi}{\mbox{\scriptsize max}}
\newcommand{\mct}{\mbox{\scriptsize MCT}}
\newcommand{\tot}{{\mbox{\scriptsize tot}}}
\newcommand{\eq}{{\mbox{\scriptsize eq}}}
\newcommand{\kb}{k_{\mbox{\scriptsize B}}}
\newcommand{\R}{{\mbox{\scriptsize R}}}
\newcommand{\RL}{\mbox{\scriptsize RL}}
\newcommand{\RS}{\mbox{\scriptsize RS}}
\newcommand{\HU}{\mbox{\scriptsize HU}}
\journalname{Eur. Phys. J. E}

\begin{document}
\title{Geometrical properties of mechanically annealed systems near
the jamming transition
}

\author{Hiromichi Matsuyama\thanksref{e1,addr1}
        \and
        Mari Toyoda\thanksref{addr1} 
        \and
        Takumi Kurahashi\thanksref{addr1} 
        \and
        Atsushi Ikeda\thanksref{addr2} 
        \and
        Takeshi Kawasaki\thanksref{addr1} 
        \and
        Kunimasa Miyazaki\thanksref{e3,addr1} 
}

\thankstext{e1}{e-mail: matsuyama@r.phys.nagoya-u.ac.jp}
\thankstext{e3}{e-mail: miyazaki@r.phys.nagoya-u.ac.jp}

\institute{Department of Physics, Nagoya University, Nagoya 464-8602, Japan\label{addr1}
\and
Graduate School of Arts and Sciences, University of Tokyo, Tokyo
153-8902, Japan\label{addr2}}

\date{Received: date / Accepted: date}

\abstractdc{
Geometrical properties of two-dimensional mixtures near the jamming transition point  
are numerically investigated using harmonic particles under mechanical training. 
The configurations generated by the quasi-static compression and
oscillatory shear deformations exhibit anomalous suppression of the
density fluctuations, known as hyperuniformity, below and above the
jamming transition.  
For the jammed system trained by compression above the transition point, 
the hyperuniformity exponent increases. 
For the system below the transition point under oscillatory shear,
the hyperuniformity exponent also increases until the shear amplitude reaches the threshold value.
The threshold value matches with the transition point from 
the point-reversible phase where the particles experience no collision
to the loop-reversible phase where the particles' displacements are non-affine
during a shear-cycle before coming back to an original position.
The results demonstrated in this paper are explained in terms of neither of universal criticality of the jamming transition nor the nonequilibrium phase transitions. 
}

\maketitle

\section{Introduction}
\label{intro}

As the density of athermal soft particles is increased, the system stops flowing and
attains rigidity at some point. 
This is called the jamming transition~\cite{Liu1998,Liu2010b,vanHecke2010}. 
In the last decade, we witnessed tremendous progress in our understanding of the jamming transition 
of frictionless particle systems, driven by the development of the
mean-field theory and extensive numerical/experimental
studies~\cite{Parisi2020book,Ozawa2017scipost,Charbonneau2017arcmp}. 
It is now well established that power-law behaviors of observables
such as the contact number, shear/bulk modulus, the radial
distributions are 
independent of spatial dimensions, as long
as the constituent particles are spherical~\cite{O'Hern2003,vanHecke2010}.
The critical behaviors are also observed in the vibrational and
rheological behavior of the
systems~\cite{Wyart2005epl,Mizuno2017pnas,Jin2018sa,Ikeda2020prl,Saitoh2020prl,Vagderg2016pre,boschan2016softmatt,Kawasaki2015pre,Otsuki2014pre}. 
Moreover, the argument is now extended to non-spherical
particles systems~\cite{Brito2018pnas}. 
These results support the universal class of the jamming transition and
bolster the success of the mean-field theory. 

There are, however, some properties that still elude the explanation
in terms of the mean-field theory and the universal criticality.
One of them is an anomalous suppression of the density fluctuations
called hyperuniformity~\cite{Torquato2003,Torquato2018pr}. 
Originally suggested in the fluctuations of the early
universe~\cite{Gabrielli2002}, hyperuniformity is observed in a variety of systems
such as avian photo-receptor cells~\cite{Jiao2014pre}, active matters~\cite{Lei2019sciadv}, 
and dilute colloidal suspensions under oscillatory shear~\cite{Tjhung2016jsmte,Weijs2015prl}.
The jammed particles system is among the first examples which exhibit hyperuniformity~\cite{Donev2005}.

Hyperuniformity is a new type of long-range correlation characterized
by the suppression of 
the static structure factor or the $q$-dependent compressibility,
$\chi(q)$, where $q$ is the wavevector. 
The system is said to be hyperuniform if $\chi(q)$ vanishes as
$q^{\alpha}$ with a positive exponent $\alpha$ instead of $\alpha=0$.
Equivalently, the hyperuniformity is defined by 
the suppression of the local density fluctuations $\sigma_\rho^2(R)\equiv \lgle \delta\rho^2(R)\rgle$, where $\delta \rho(R)$
is the density fluctuations inside an observation window of the radius $R$~\cite{Torquato2018pr}.
For hyperuniform systems, $\sigma_{\rho}^2(R)$ decreases as $R^{-\beta}$ 
$(\beta >d)$ where $d$ is the spatial dimension.
This is faster than $R^{-d}$ for equilibrium liquids.
The exponent $\beta$ is related to $\alpha$ by $\beta = d + \alpha$, if
$\alpha \leq 1$~\cite{Torquato2018pr}. 
Hyperuniformity was presumed to be another sign of the jamming criticality but
the power-law behavior of $\chi(q)\sim q^{\alpha}$ is observed only in the finite range of the
wavevector windows~\cite{Ikeda2015pre,Ikeda2017pre}. 
It is argued that $\chi(q)$ becomes strictly zero only when the jamming configuration is
prepared from the ideally close-packed amorphous state (or the ideal glass
state)~\cite{Godfrey2018prl}. 
However, a numerical study on the jamming configurations generated by
quenching thermally annealed supercooled fluids shows that hyperuniformity is not
enhanced~\cite{Ozawa2017scipost}.
Distinct but related jammed system of confluent living tissues also
show hyperuniformity, but it has been discussed that hyperuniformity is not
directly related to the fluid-solid transition~\cite{Zheng2020sm}.
Recently, hyperuniformity was also observed 
in the inherent structures
of supercooled liquids~\cite{Mitra2021Jstat} and in the frictional jammed systems~\cite{Yuan2021prr}.

In this paper, we numerically study the geometrical properties, with a
special emphasis on hyperuniformity, of particle configurations near and
below the jamming transition point under several mechanical training protocols. 
One protocol is the isotropic compression in which an unjammed
configuration is compressed quasi-statically and isotropically up to a
maximal density $\varphi_{\MAX}$ above the jamming transition point and then decompressed
again until the pressure or the energy vanishes~\cite{Kumar2016granular}.  
We define the jamming transition point $\varphi_\mathrm{J0}$, as the value at which the pressure/energy becomes finite
on the way of compression. 
As reported in Ref.~\cite{Kumar2016granular}, 
the jamming density at which the pressure/energy vanishes during the
decompression, which we denote as $\varphi_\mathrm{J}$, is larger than
$\varphi_\mathrm{J0}$ and depends mildly on the amplitude of the compression cycle $\varphi_{\MAX}$. 
This means that the jammed configuration is trained and the system finds
more efficient packing during this mechanical cycle, much the same way
as the jamming density increases as the parent fluid before quenching to generate the jamming configuration is equilibrated at
lower temperatures~\cite{Chaudhuri2010,Ozawa2012prl,Ozawa2017scipost}.
For this reason, we use the word ``training'' for the same meaning as ``annealing'' in this paper.
We investigate the density fluctuations of the jammed configurations at
$\varphi_\mathrm{J}$ after one compression-decompression cycle.

Another system we study is the configurations slightly below
$\varphi_\mathrm{J0}$ which is subject to quasi-static oscillatory shear. 
For a given configuration, we apply a quasi-static shear deformation cycle with a small amplitude.  
The cycles are applied until the system reaches the steady state
and then we monitor the density fluctuations. 
The Poissonian-like large density fluctuations are suppressed
quickly as the training shear amplitudes increases.
In both systems, we find that 
the mechanical training enhances the
hyperuniformity of the system.

This paper is organized as follows.
After describing the method and the detail of the system which we study in Section 2, 
we first revisit hyperuniformity of a two-dimensional jammed configuration generated
in the vicinity of the jamming transition point of the poorly trained system in Section 3.
In Section 4, we analyze the rattler populations and hyperuniformity for
the system generated by the compression training. 
We analyze hyperuniformity for the unjammed configurations under
oscillatory shear in Section 5. 
We conclude this paper in Section 6.

\section{Method and Model}
\label{sec:2}

The system we study is a two-dimensional equimolar
binary mixture of frictionless particles with diameters $\sigma_L$
and $\sigma_S$ placed in a square box with periodic boundary conditions. 
The size ratio of small and large particles is $\sigma_L/\sigma_S= 1.4$. 
The particles interact with the harmonic potential 
defined by 
\begin{equation}
 U(r_{ij}) = \frac{\varepsilon}{2}\left\{ 1 - (r_{ij}/\sigma_{ij})\right\}^2,
\end{equation}
if the inter-particle distance of the $i$-th and $j$-th particles is
$r_{ij} <\sigma_{ij}$ and 0 otherwise. 
Here $\sigma_{ij} = (\sigma_i + \sigma_j)/2$ is the average diameter of the $i$-th and $j$-th 
particles.
We use $\sigma_S$, $\varepsilon$,  and $\varepsilon/k_B$ as the units of
length, energy, and temperature, respectively. 

For all simulations presented in this paper, we first generate jammed or
unjammed configuration by quenching the random configuration at
$T=\infty$ to $T=0$ at a target volume fraction $\varphi$. 
For both mechanical compression (Section \ref{sec:4}) and the
oscillatory shear (Section \ref{sec:5}), the configuration is optimized
using the FIRE algorithm~\cite{Bitzek2006prl} for every infinitesimal
step of the deformation. 
The number of samples is from 300 to 900.
The details of the numerical method for the different protocols are given
in each section. 

\section{Hyperuniformity of the poorly trained system}
\label{sec:3}

In this section, we revisit hyperuniformity of the poorly trained jammed configuration. 
Hyperuniformity in the vicinity of the jamming transition point $\varphi_\mathrm{J}$ is
well documented for both two~\cite{Wu2015pre,Chieco2018pre} 
and three dimensional
systems~\cite{Donev2005,Berthier2011c,Ikeda2015pre,Ikeda2017pre,Ozawa2017scipost}, 
but the properties of the hyperuniform configurations in two-dimensional systems have not been fully explored. 

We generate the jamming configuration using the same method as in previous studies~\cite{Chaudhuri2010,Kawasaki2020}.
The initial random particle configuration prepared at $\varphi_\mathrm{ini}=0.835$ is 
quasi-statically compressed by an incremental small density $\Delta\varphi= 10^{-5}$ 
up to the maximum density $\varphi_{\MAX} = 0.845$.
At every compression step, the system is stabilized using the FIRE algorithm~\cite{Bitzek2006prl}. 
When the density reaches $\varphi_{\MAX}$, the system is decompressed with the step of $\Delta\varphi$.
We monitor the energy $E$ at every step and we judge the system is unjammed if the energy per one particle $e=E/N$ becomes smaller than $e=10^{-16}$.
We determine the jamming transition point $\varphi_\mathrm{J}$ as the
first density at which the system is unjammed during the decompression steps. 
We obtain the mean value of the jamming transition point 
$\varphi_\mathrm{J} = 0.842$.
In the analysis below, we monitor the density fluctuations at the jamming transition point of each sample. 
Throughout this section, the system size is $N=3000$ and we take the
ensemble average over the 300 samples.

We use the wavevector-dependent compressibility $\chi(q)$ to monitor the density
fluctuations. 
For binary mixtures, $\chi(\bm{q})$ is defined by 
\begin{equation}
 \chi(\bm{q}) 
= \frac{S_{SS}(\bm{q})S_{LL}(\bm{q})- S_{LS}^2(\bm{q})}{x_S^2S_{LL}(\bm{q})+ x_{L}^2S_{SS}(\bm{q})- 2x_{S}x_{L}S_{LS}(\bm{q})},
\label{eq:chiq}
\end{equation}
where $x_S=x_L=1/2$ is the molar fraction of the small and large
particles,  
$S_{\nu\mu}(\bm{q}) = \lgle \delta\rho_{\nu}(\bm{q})\delta\rho_{\mu}^{\ast}(\bm{q})\rgle/N$
($\nu, \mu= S, L$) is the static structure factor matrix for the
binary mixture~\cite{Berthier2011c}. 
Note that the small wavevector limit of $\chi(q)$ is
related to an isothermal compressibility by $\chi(q\rightarrow 0) =
\rho\kb T\chi_T$ in equilibrium.
It is established that $\chi(q)$ is a better observable to detect the
hyperuniformity than the total structure factor $S(q)
=\sum_{\nu\mu= S,L}S_{\nu\mu}(q)$~\cite{Berthier2011c}
\footnote{In the ensemble-averaging of $\chi(q)$, we expand the system slightly to fix the simulation box-size for every sample. 
This is necessary since each sample jams at different volume fractions.
We carefully checked that the effect of the small variation of the
system size is negligible.}. 
Recently, several variants of $\chi(q)$ have been
proposed~\cite{Wu2015pre}, such as the local volume fraction correlation
functions.  
The wavevector windows where hyperuniformity is observed vary
depending on the variants  but the long-length (small $q$) behavior is 
insensitive to the definition~\cite{Wu2015pre}.

\begin{figure}[htb]
\includegraphics[width=1.0\columnwidth,angle=-0]{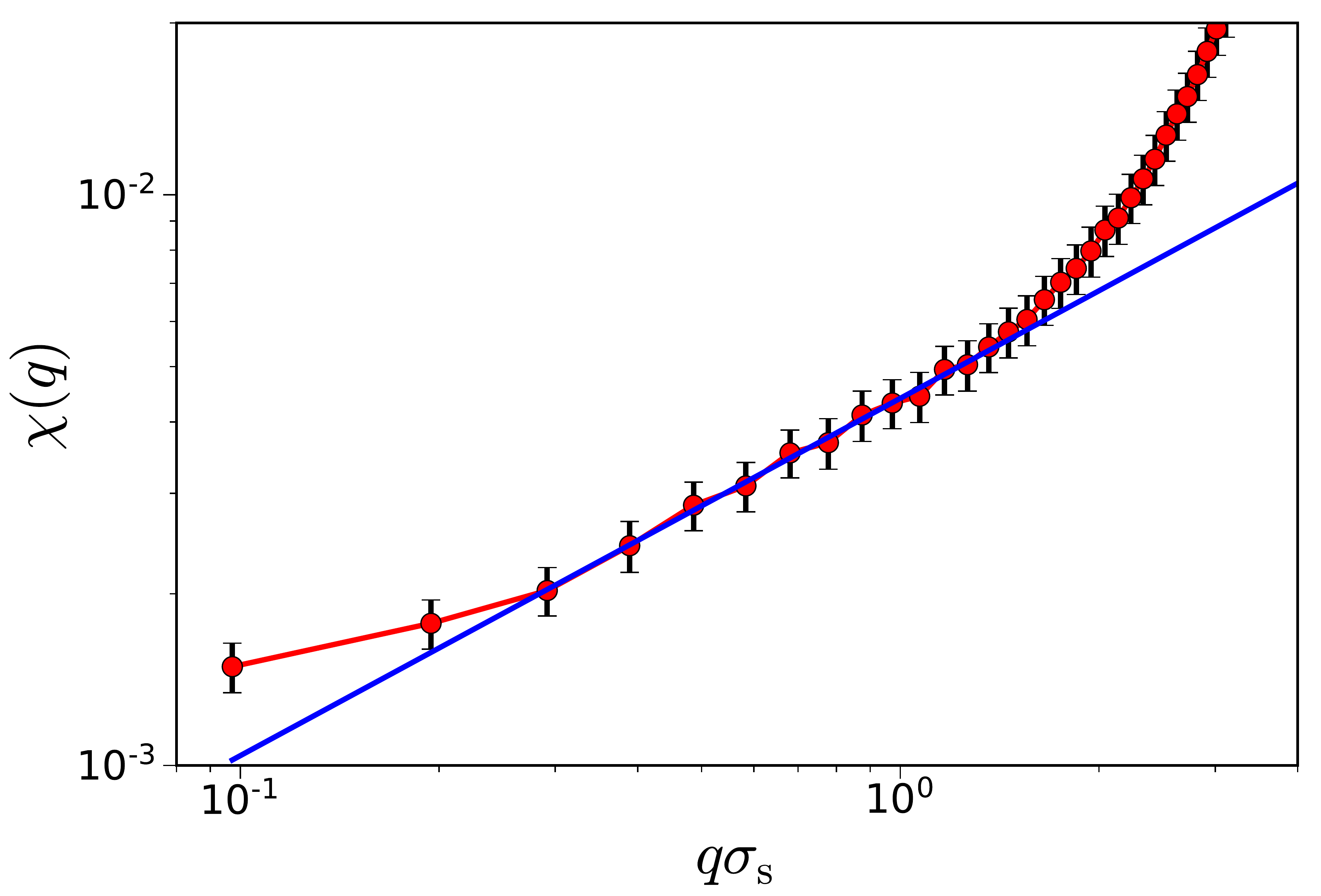}
\caption{
$\chi(q)$ at $\varphi_\mathrm{J}$ for the poorly trained system. 
The red circles are simulation results with standard errors. 
The blue solid line is the power-law fit by $\chi(q)\sim q^{\alpha}$ with the exponent $\alpha = 0.626 \pm 0.012$.
The fitting was made in the window of $0.3 \lesssim q  \lesssim 1.3$.
The value of $\alpha$ is smaller than $\alpha \approx 1 $ which is reported in previous study in three dimensional system.
}
\label{fig:1}
\end{figure}
\begin{figure}[htb]
\includegraphics[width=1.0\columnwidth,angle=-0]{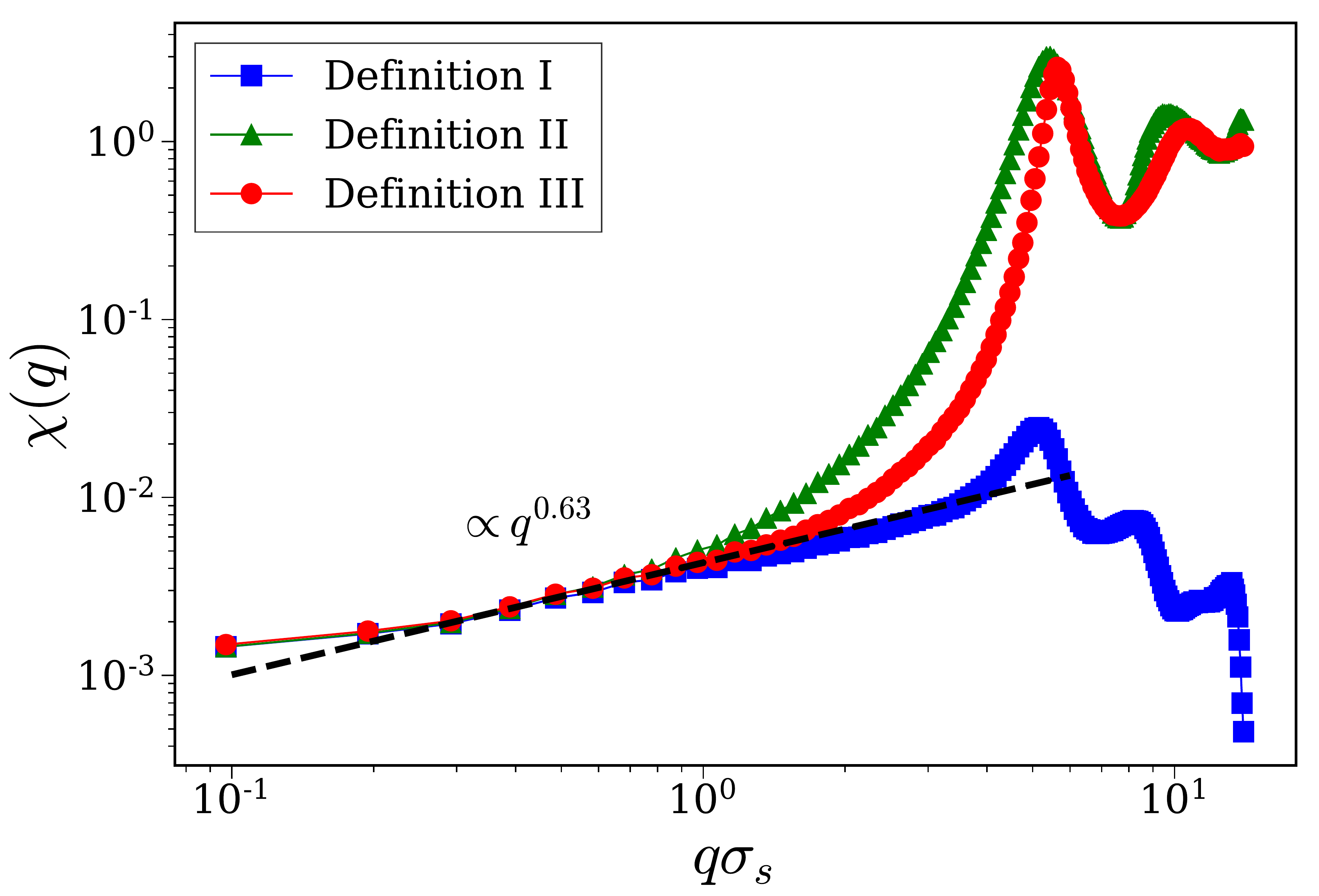}
\caption{
$\chi(q)$ computed using three different definitions (see the
 text). Definition III is identical with that shown in Figure \ref{fig:1}.
 The line  of $q^{0.63}$ is guide for eyes.
}
\label{fig:chi_def}
\end{figure}

In Figure \ref{fig:1}, we show $\chi(q)$ at $\varphi_\mathrm{J}$.
$\chi(q)$ develops the power law region in
the intermediate window of $0.3 \lesssim q  \lesssim 1.3$, 
as it is the case for the three dimensional systems~\cite{Donev2005,Berthier2011c,Ikeda2015pre,Ikeda2017pre,Ozawa2017scipost}. 
This algebraic behavior is not observed  in the small-$q$ limit and
$\chi(q)$ bends up around $q_{\HU} \lesssim 0.3$.
This means that the system is not rigorously hyperuniform and 
the spatial correlation persists only up to a length
scale $q_{\HU}^{-1}$. 
We shall refer to $q_{\HU}^{-1}$ as the hyperuniform length.
In the intermediate region, 
$\chi(q)$ is fitted well by $\chi(q) \sim q^{\alpha}$ 
with an exponent $\alpha = 0.626 \pm 0.012$, where we use the least square for fitting.
This exponent is far smaller than $\alpha\approx 1$ for the three-dimensional systems. 
It may sound surprising because $\alpha \approx 1$ were univocally reported in 
the literature on hyperuniformity of the jammed systems, irrespective
of the spatial dimension.  
However, this result is not new. 
Although it has not been explicitly acknowledged in the previous
studies~\cite{Wu2015pre,Dreyfus2015pre,Chieco2018pre}, close inspection of the data shows that the values of exponents
reported for the two-dimensional systems are consistent with ours.
We check that the value of $\alpha$ does not depend on the
definition of $\chi(q)$.  
Several types of $\chi(q)$ defined by different metrics
have been proposed and studied recently~\cite{Berthier2011c,Wu2015pre,Hexner2018prl}.  
Here we examine three definitions of $\chi(q)$'s introduced in Ref.~\cite{Wu2015pre}. 
Definition I~\cite{Zachary2011prl}: $\chi_{\mbox{\scriptsize I}}(q) \sim \lgle |\Delta(q)|^2\rgle$, 
where
$\Delta(q)$ is the Fourier transformation of the local volume fraction defined by 
$\sum_i \Delta_i(\bm{r} - \bm{r}_i)$ where 
$\Delta_i$ is the step function which is 1 if $\bm{r}$ is sitting inside
the $i$-th disc at $\bm{r}_i$ and 0 otherwise. 
Definition II~\cite{Berthier2011c}: $\chi_{\mbox{\scriptsize II}}(q) \sim \lgle |\delta\varphi_(q)|^2\rgle$,
where $\delta\varphi(q)$ is the Fourier transformation 
$\varphi(\bm{r}) = \sum_i v_i \delta_i(\bm{r} - \bm{r}_i) $
where $v_i$ is of the area of the $i$-th disc. 
Definition III~\cite{Berthier2011c} is the $q$-dependent
compressibility defined by \eqref{eq:chiq}. 
We show these three $\chi(q)$'s in Figure~\ref{fig:chi_def}.
Three $\chi(q)$ match in the small $q$ region with the identical
power-law exponent.  
Although the width of window over which the power-law is observed varies
depending on the definition and the shape at large $q$'s are salient,   
it is evident that the hyperuniform behaviors are robust irrespective of
the choice of metrics. 

The smaller exponent of $\alpha \approx 0.6$ may be understood
by recalling that the upper critical dimension of the jamming transition
is 2. 
Let us consider an ideally hyperuniform jammed configuration. 
Then, 
the fluctuations $\sigma_\phi^2(R)\equiv\lgle \delta\phi^2\rgle$ of a
subsystem of the size $R$ should be scaled as $R^{-d-\alpha}$ in the
large $R$ limit and its Fouier representation 
should behave as $\chi_\phi(q)\sim q^{\alpha}$~\cite{Torquato2018pr}. 
Here $\phi$ is a metric such as number density.
On the other hand, the mean-field theory claims that 
$\sigma_\phi^2(R) \sim N^{-\omega}$ should be scaled by $N$ rather
than $R$ in the large $N$ (or $R$) limit~\cite{Binder1985prb,Hexner2019prl},
where $\omega$ is a dimension-independent exponent. 
Comparing the two expressions, one concludes $d+\alpha= d\omega$,
or $\alpha \propto d$. 
Therefore, if $\alpha=1$ for $d=3$ as it is established, 
we should expect $\alpha=2/3\approx 0.6$ for $d=2$.
Of course, this argument applies in the large $N$ limit and in the
ideally hyperuniform system.
In the finite system, the scaling behavior of the real-space representation of the fluctuations $\sigma_\phi^2(R)$ is 
polluted by, for example, the shape effect of
subsystems~\cite{Ikeda2017pre} and, in realistic jammed systems, 
$\chi_\phi(q)\sim q^{\alpha}$ breaks down below $q_{\HU}$.  
However, $\chi_\phi(q)\sim q^{\alpha}$ would survive in the intermediate
window $q > q_{\HU}$ and should be less sensitive to the finite size effect.

The exponent of $\alpha\approx 0.63$ are also observed in
the system under nonequilibrium setting as we shall demonstrate in the
following sections.

\section{Mechanically trained systems by the compression and decompression cycle}
\label{sec:4}

In this section, we investigate the jammed system under mechanical
training by applying the compression-decompression cycle to the system. 
Kumar {\it et al.} has shown that, if a system jammed at a value of $\varphi_\mathrm{J0}$
is compressed by a certain amount up to $\varphi_{\MAX}$ and then decompressed, the system is
unjammed at a value $\varphi_\mathrm{J}$ which is slightly larger than $\varphi_\mathrm{J0}$~\cite{Kumar2016granular}. 
If the compression-decompression cycle
is applied multiple times,
$\varphi_\mathrm{J}$ increases more.  
In other words, the jamming transition point $\varphi_\mathrm{J}$ shifts to a larger value
as the system is trained mechanically by compression. 
This is analogous to the thermal annealing effect in which 
$\varphi_\mathrm{J}$ increases if the parent fluid before quenching to
generate the jammed packing is equilibrated at lower temperatures~\cite{Chaudhuri2010,Ozawa2012prl}. 
Here we investigate how the geometrical properties of the jammed
configurations change by such mechanical training.
For thermal annealed systems for which the system parameters are
carefully calibrated to facilitate the equilibration at high densities,
it has been shown that (i) the number of rattlers increases
substantially and (ii) concomitantly, hyperuniformity is mitigated as
the system is more annealed~\cite{Ozawa2017scipost}. 
It is somewhat counter-intuitive as we would naively expect that more
stabilized and optimized amorphous configurations by annealing would
prefer the fewer particles with no contacts and also would exhibit stronger
hyperuniformity as the configurations are closer to those of the ideal glass~\cite{Godfrey2018prl}.

\begin{figure}[htb]
\includegraphics[width=1.0\columnwidth,angle=-0]{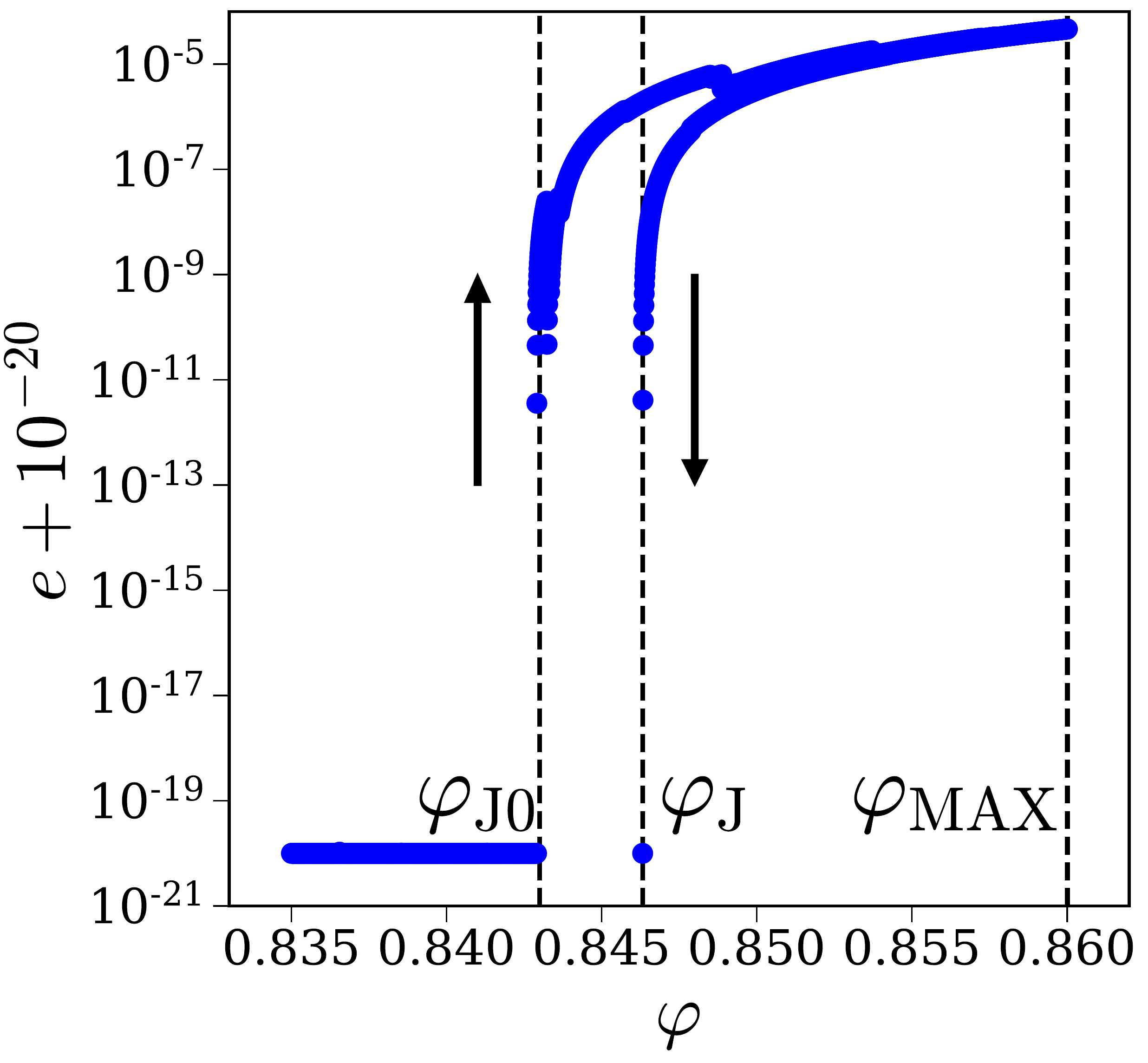}
\caption{The change of the energy during one compression-decompression
 cycle. The line is shifted by $10^{-20}$ upward to show the unjammed state points where $E=0$. 
$E$ abruptly increasing and it defines $\varphi_\mathrm{J0}$ of the poorly trained system and $E$ vanishes at slightly higher density $\varphi_\mathrm{J}$ on the way back from the training cycle. The position of $\varphi_{\MAX}$ is also shown.}
\label{Fig_compression.jpg}  
\end{figure}

We train the system mechanically using a quasi-static volume compression-decompression
for one cycle using the protocol employed by Kumar {\it et al.}~\cite{Kumar2016granular}.  
As explained in the previous section, the initial configuration prepared at
$\varphi_\mathrm{ini}=0.835$ slightly below $\varphi_\mathrm{J}$ is quasi-statically compressed to a maximum
density $\varphi_{\MAX}$ with the step size of $\Delta\varphi= 10^{-5}$.  
As the density reaches $\varphi_{\MAX}$, the system is decompressed with
the same step size.  
We identify $\varphi_\mathrm{J}$ as the point where the energy per one particle $e=E/N$
becomes smaller than $e=10^{-16}$.
For the analysis of the rattlers, we use the system size $N=1000$ and
average over 600 samples. 
For the analysis of hyperuniformity, $N=3000$ and the number of samples is 300. 
The typical behavior of $e$ during this compression cycle is shown in Figure \ref{Fig_compression.jpg}.  
We simulate various values of $\varphi_{\MAX}$ ranging
from 0.845 up to 1.3.

\subsection{Rattlers}

Before discussing hyperuniformity, we comment on the change in the number of rattlers under mechanical training. 
It is documented that the jammed configurations generated from thermally annealed fluid which is equilibrated at a temperature well below the so-called onset temperature of the supercooled state tend to generate
more rattlers than the jammed configurations generated by fast quench from
the high-temperature fluids (or random configuration)~\cite{Ozawa2017scipost}. 
We investigate if this counter-intuitive result holds for the jammed configurations
under the mechanical training. 

\begin{figure}[htb]
\includegraphics[width=1.0\columnwidth,angle=-0]{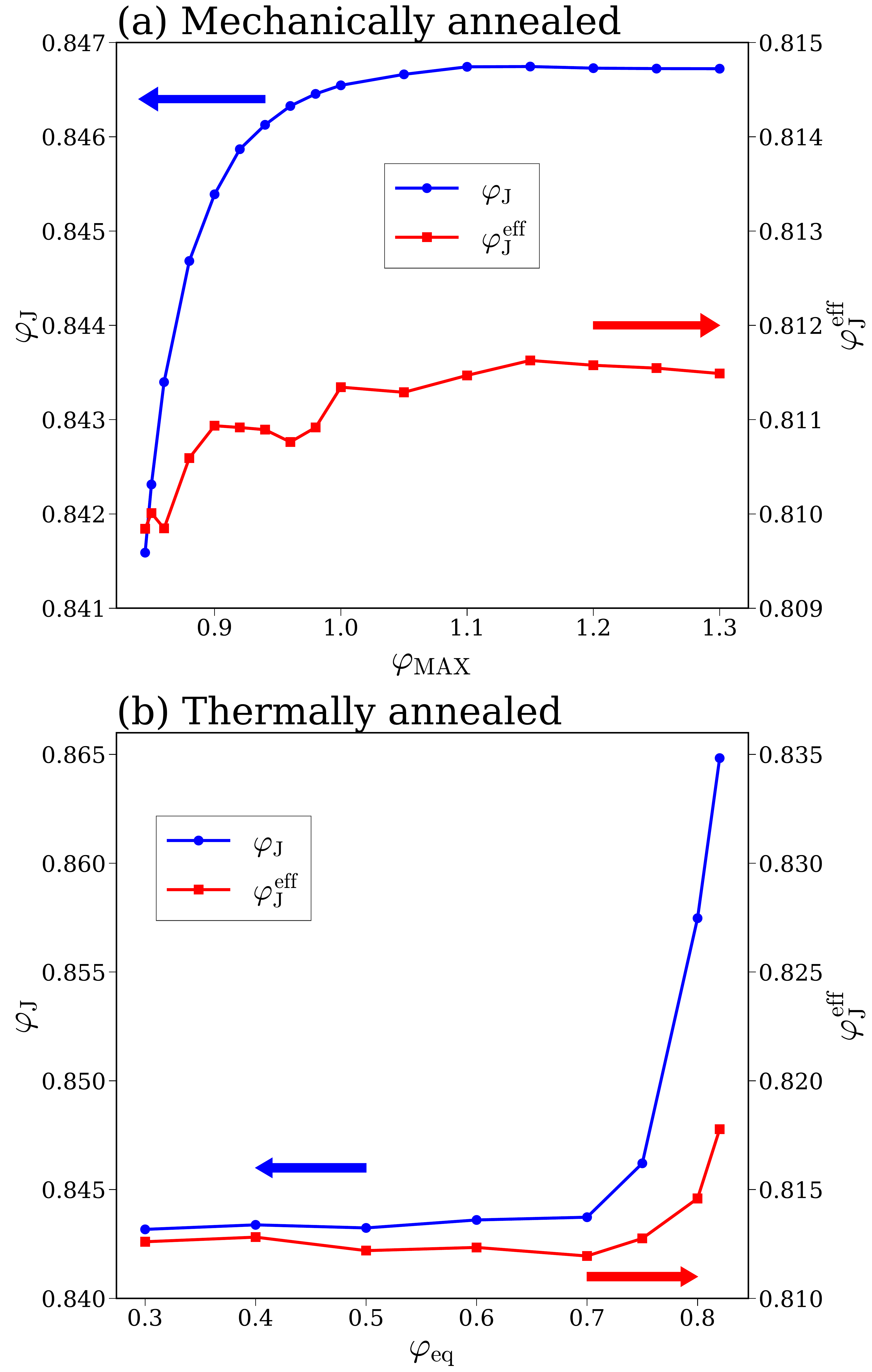}
\caption{(a) The jamming transition point $\varphi_\mathrm{J}$ (the blue circles) 
and the effective jammed density $\varphi_\mathrm{J}^{\eff}$ (the red squares)
as a function of $\varphi_{\MAX}$.  
$\varphi_\mathrm{J}^{\eff}$ is the packing fraction calculated for the particles that contribute to maintain the structure.
(b) The same results for the $22$\% polydisperse harmonic spheres
 deeply thermally annealed. $\varphi_{\eq}$ is the volume fraction of
 the parent fluid at a finite $T= 1.0\times 10^{-6}$ before quench to
 generate the jamming configuration. $\varphi_\mathrm{J}$ is constant up to
 $\varphi_{\eq}\approx 0.7$ and then sharply spike about 3\% at
 $\varphi_{\eff}= 0.82$.
 The scales of each axis of the two-axis plot (a) and (b) are aligned.
}
\label{Fig_phiJ-vs-phieff.jpg}
\end{figure}

\begin{figure}[htb]
\includegraphics[width=1.0\columnwidth,angle=-0]{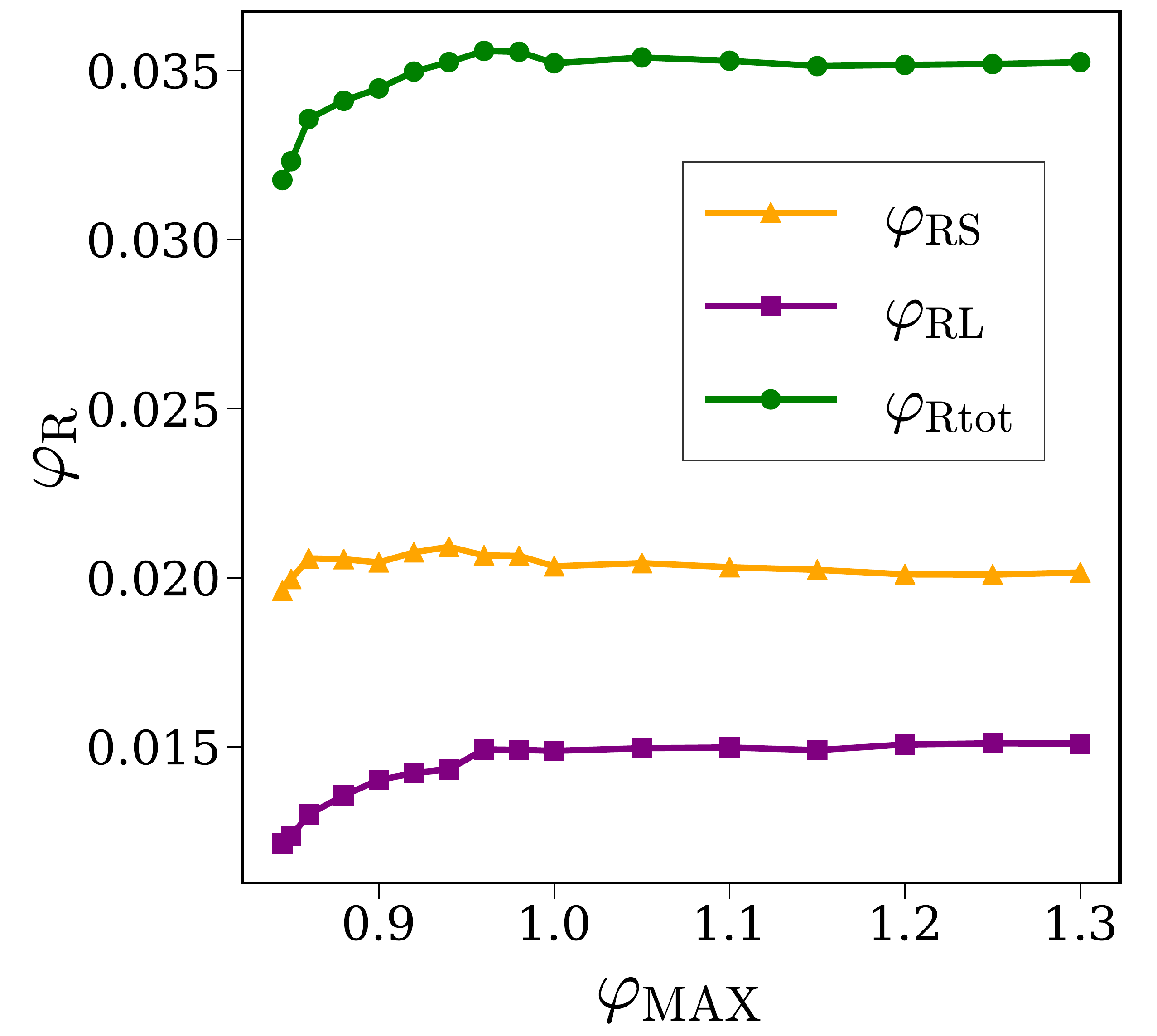}
\caption{The dependence of the volume fraction of total rattler
 $\varphi_{\R}$ (green circles), small rattlers $\varphi_{\RS}$ (yellow triangles), 
and large rattlers $\varphi_{\RL}$ (purple square) on the maximal
 compression $\varphi_{\MAX}$.
 The amplitude of increase of $\varphi_{\R}$ on $\varphi_{\MAX}$ is larger than $\varphi_{\RS}$.
 Thus, increase of $\varphi_{\R}$ is dominant contribution of $\varphi_{\R\tot}$.}
\label{Fig_phiL-vs-phiS.jpg}
\end{figure}

As reported in Refs.~\cite{Kumar2016granular,Kawasaki2020}, $\varphi_\mathrm{J}$ 
is a function of $\varphi_{\MAX}$  and increases slowly with $\varphi_{\MAX}$ (see Figure \ref{Fig_phiJ-vs-phieff.jpg} (a)).
$\varphi_\mathrm{J}$ sharply increases up to
$\varphi_{\MAX}=1$ and then saturate at slightly below 
$0.847$ (the blue filled circles). 
The maximal increase from $\varphi_\mathrm{J}$ for poorly annealed system is about 0.61\%, which is
comparable with the value reported for the three dimensional system~\cite{Kumar2016granular}.  
We then count the number of rattlers in each jammed configuration.
We defined the rattler by the particle whose contact number $Z$ is $0$. 
The green circles in Figure~\ref{Fig_phiL-vs-phiS.jpg} show the
total volume fraction of the rattlers $\varphi_{\R \tot}$ as a function of $\varphi_{\MAX}$.
It demonstrates that the ratters increase as the system is
mechanically trained. 
To quantify the contribution of the particles which maintain the 
jammed configurations, we define the effective jammed density 
$\varphi_\mathrm{J}^{\eff}= \varphi_\mathrm{J} - \varphi_{\R}$ which is plotted in Figure
\ref{Fig_phiJ-vs-phieff.jpg} (red square). 
The result shows that the increase of $\varphi_\mathrm{J}^{\eff}$ with $\varphi_{\MAX}$
is far smaller compared with that of $\varphi_\mathrm{J}$.
Compared with 0.61\% for $\varphi_\mathrm{J}$, $\varphi_\mathrm{J}^{\eff}$ increased only
by 0.22\%, which implies that an increase of $\varphi_\mathrm{J}$ 
is dominated by the increase of the rattlers.
In other literature, the rattler is defined by the particle whose
contact number satisfies $Z <d+1$~\cite{Ozawa2017scipost}. 
With this definition, $\varphi_\mathrm{J}^{\eff}$ shifts up by about $0.002$ but
the qualitative behavior is not altered.

We also compare this result with that of a thermally annealed system. 
We adopt the polydisperse system because it is much easier to
equilibrate the system at very high densities using an efficient swap
Monte Carlo simulation, than that of binary systems~\cite{Ninarello2017prx}.
We use the 22\% polydisperse systems with the diameter distribution 
given by $P(\sigma) \propto \sigma^{-4}$.
The ratio of minimum and the maximum diameter is given by $\sigma_{\mini}/\sigma_{\maxi}=0.45$~\cite{Liao2019prx}. 
We equilibrate the system of $N=1024$ for various densities 
$\varphi_{\eq}$ at a finite temperature $T_{\eq}=1.0\times10^{-6}$ and then quench the system to $T=0$. 
The results are insensitive to $T_{\eq}$ as long as it is low enough and $\varphi_{\eq}$ is not
large, where the thermodynamic properties of harmonic discs are well
approximated by those of hard discs. 
In Figure \ref{Fig_phiJ-vs-phieff.jpg} (b), we show $\varphi_\mathrm{J}$ (blue
circles) and $\varphi_\mathrm{J}^{\eff}$ (red squares). 
As the equilibrium density $\varphi_{\eq}$ increases, 
$\varphi_\mathrm{J}$ bents upward sharply $\varphi_{\eq} \approx 0.7$. 
The position of the bent in our data is close to the mode-coupling transition point $\varphi_{\mct}=0.795$
reported in Ref.~\cite{Liao2019prx} for a system similar to ours (except
for a different polydispersity).
The increase of $\varphi_\mathrm{J}^{\eff}=\varphi_\mathrm{J} - \varphi_{\R}$ synchronizes with that of
$\varphi_\mathrm{J}$ but the degree of the increase is far milder than
$\varphi_\mathrm{J}$, similarly with the result of compression training. 
This means that the rattlers are primarily responsible for the
increase of the jamming packing. 
This is somewhat surprising because the denser amorphous packing 
with both thermal annealing and mechanical training is the consequence
of the system finds a more stable configuration by descending to the
lower entropy landscapes marked by the mode-coupling crossover, which is
described by the mean-field picture, whereas the rattlers are objects
which is not prescribed by the mean-field scenario. 

More surprising is that the large rattlers predominantly contributes to the
increase of the jamming density. 
Figure~\ref{Fig_phiL-vs-phiS.jpg} shows the dependence of the volume
fractions of total rattlers $\varphi_{\R\tot}$, the rattlers of small discs
$\varphi_{\RS}$, and of large discs $\varphi_{\RL}$ on $\varphi_{\MAX}$
for the mechanical trained binary mixture. 
Note that the increase of the rattler volume fraction $\varphi_{\R}$ is
dominated by the increase of the large rattlers $\varphi_{\RL}$ and the
number of the small rattlers remains almost constant. 
\begin{figure}[htb]
\includegraphics[width=1.0\columnwidth,angle=-0]{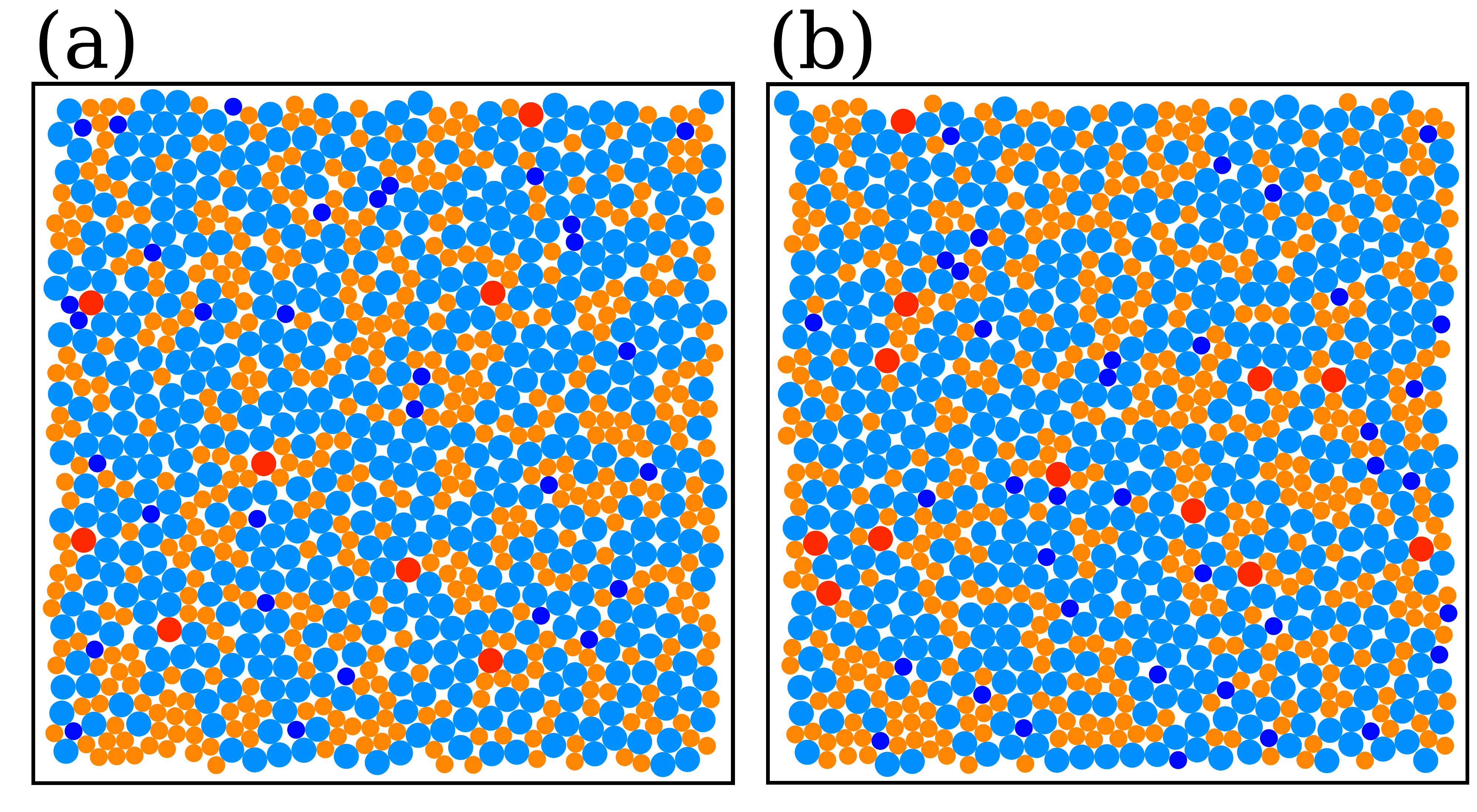}
\caption{The snapshot of the jammed configurations for the poorly trained system (a) 
and for the trained system with $\varphi_{\MAX}=0.98$ (b).
 The Jamming transition points of these samples are 
 $\varphi_\mathrm{J}=0.840$ for the poorly trained system and 
  $\varphi_\mathrm{J}=0.846$ for the trained system with $\varphi_{\MAX}=0.98$.
 The small rattlers (the blue disks) and the large rattlers
 (the red disks) are randomly embedded in the non-rattlers of the small
 (orange disks) and large (light blue disks) disks but the change
 is barely discernible.} 
\label{Fig_snapshot.png}
\end{figure}
Figure~\ref{Fig_snapshot.png} is the snapshot of the
jammed configurations of (a) poorly trained system and (b) mechanically trained
system.  
Except for the slight decrease of free-area (blank regions) for the denser packing in (b), it is difficult to spot any qualitative changes due to the increase of the large rattlers. 
It is not clear how the large rattlers contribute to the efficiency of the packing.
We need more quantitative analysis, such as the spatial correlation of the
rattlers for a larger system size to dissolve this conundrum~\cite{Atkinson2013pre,Ozawa2017scipost}.

\subsection{Hyperuniformity}

Next we investigate hyperuniformity of the mechanically trained
system by compression at the jamming transition point $\varphi_\mathrm{J}$. 
We use the $q$-dependent compressibility $\chi(q)$ defined by
Eq.~\eqref{eq:chiq} to monitor hyperuniformity. 
\begin{figure}[htb]
\includegraphics[width=1.0\columnwidth,angle=-0]{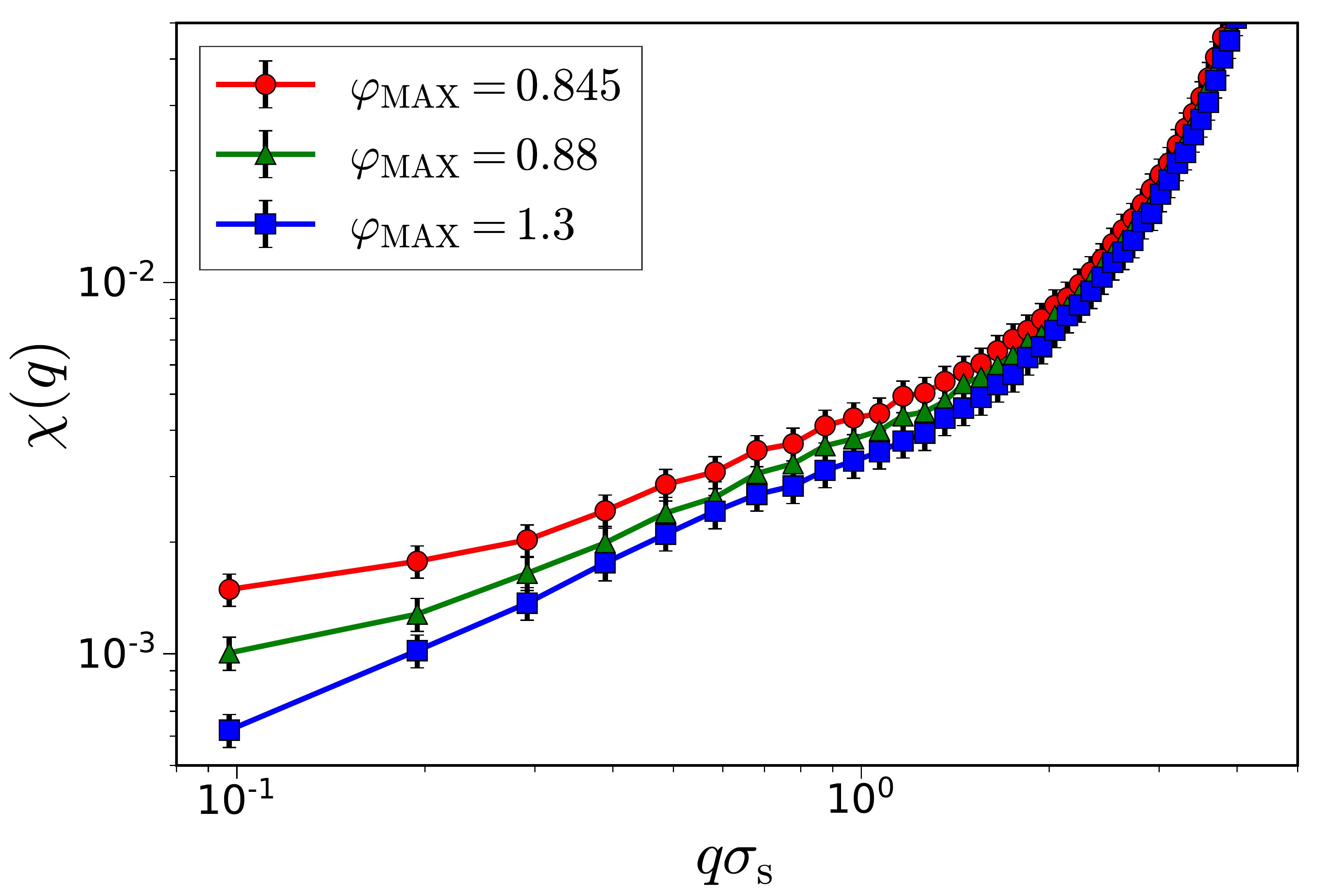}
\caption{$\chi(q)$ for various $\varphi_{\MAX}$. $q$ is scaled by
 $\sigma_\mathrm{S}$. 
 The algebraic decay becomes stronger with increasing the strength of mechanical annealing $\varphi_{\MAX}$.
 Moreover, the intermediate window which emerges hyperuniformity moves lower $q$ with increasing $\varphi_{\MAX}$.
 }
\label{Fig_Sq_Compressed.png}
\end{figure}

\begin{figure}[tbh]
\includegraphics[width=1.0\columnwidth,angle=-0]{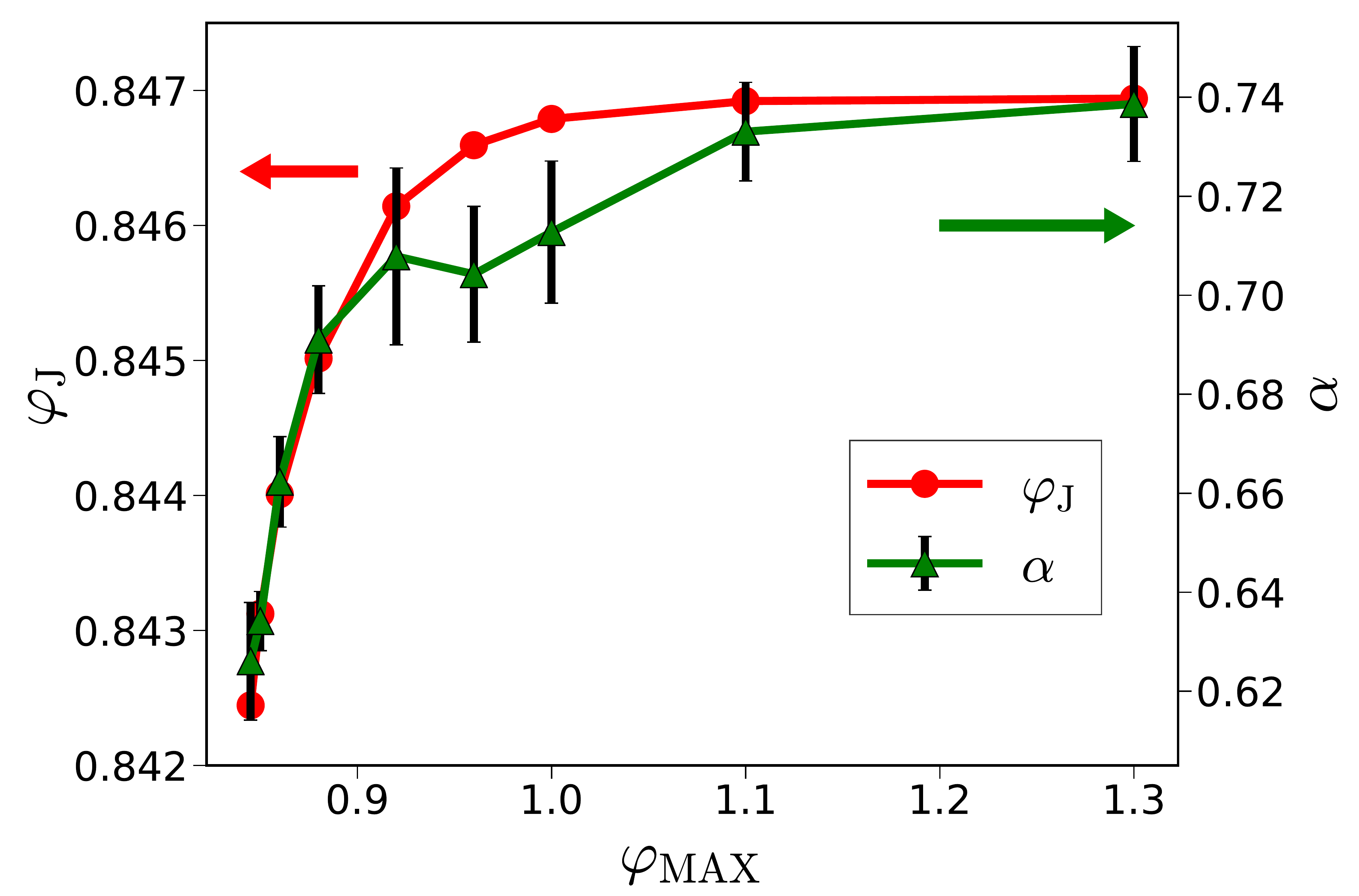}
\caption{The dependence of $\alpha$ on $\varphi_{\MAX}$ and $\varphi_\mathrm{J}$.
$\alpha$ sharply increases as increasing $\varphi_{\MAX}$ and reaches to 
$\alpha\approx 0.74$ at $\varphi_{\MAX}\approx
1.3$.
This increase of $\alpha$ on $\varphi_{\MAX}$ shows similar behavior of $\varphi_\mathrm{J}$. 
}
\label{Fig_alpha_Compressed.png}
\end{figure}

We evaluate $\chi(q)$ for jammed configurations generated after
one cycle of compression-decompression with various $\varphi_{\MAX}$'s.
We take the ensemble average of over 300 samples for each $\varphi_{\MAX}$. 
The $q$-dependence of $\chi(q)$ is shown in Figure~\ref{Fig_Sq_Compressed.png}. 
For all $\varphi_{\MAX}$'s, we observe the algebraic behavior of
$\chi(q) \sim q^{\alpha}$ at the intermediate window of $0.3 \lesssim q  \lesssim 1.3$. 
The result of least annealed system ($\varphi_{\MAX}$=0.845) is
identical to Figure~\ref{fig:1} for which 
the hyperuniformity exponent is $\alpha \approx 0.63$.
For larger $\varphi_{\MAX}$'s, one observes slight but distinct changes
in the behavior of $\chi(q)$; 
the hyperuniformity exponent $\alpha$ systematically increases. 
We plot the dependence of $\alpha$ on $\varphi_{\MAX}$ 
in Figure~\ref{Fig_alpha_Compressed.png} (green triangles).
$\varphi_\mathrm{J}$ (red circles) are plotted on the same figure to
highlight the similarity. 
The exponent $\alpha$ is obtained by fitting in the region where the algebraic
behaviors are observed.
The fitting windows are 
$0.3 \lesssim q  \lesssim 1.3$ for $0.845 \lesssim \varphi_{\MAX}  \lesssim
0.9$ and 
$0.1 \lesssim q  \lesssim 1.0$ for $0.9 < \varphi_{\MAX}$. 
We observe that the increase of $\alpha$ is similar to that of
$\varphi_\mathrm{J}$, except for a small decrease around
$\varphi_{\MAX}\approx 0.95$. 
At $\varphi_{\MAX}\approx 0.95$, 
we observed the effective jammed density $\varphi_\mathrm{J}^{\eff}$ decreases 
in Figure~\ref{Fig_phiJ-vs-phieff.jpg}. 
This hints a correlation between the number of rattler particles and the hyperuniform
scaling and worth further investigations.
The maximal value of $\alpha$ at the largest $\varphi_{\MAX}$ we studied
is $\alpha\approx 0.739 \pm 0.012$.
The increase of $\alpha$ implies the suppression of the density fluctuations of $\lgle \delta\rho^2(R)\rgle$.  
If the argument given in Section~\ref{sec:3} holds, 
the system size dependence of the width of the distributions of the jamming density
should become stronger for the mechanically trained system. 
It is natural to expect that the increase of $\alpha$ should be observed
also for thermally annealed systems. 
In a recent study, Chieco {\it et al.}~\cite{Chieco2018pre} has
investigated hyperuniformity of a two-dimensional mixture for jammed
packing generated at very low temperatures. 
Though it is not explicitly mentioned therein, 
their results hint that the exponent $\alpha$ faintly larger than those of quickly quenched systems, 
which is similar to our observation. 
The fact that $\alpha$ reaches 0.75 and does not increase further with
$\varphi_{\MAX}$ implies that there exists the upper limit in the effect of the mechanical training. 
As it is established in the study of nonlinear rheology of the
oscillatory sheared systems~\cite{Parmar2019prx,Yeh2020prl,Bhaumike2021pnas},  
mechanical training plays a similar role as thermal annealing.
The system descends the energy landscapes to reach stable configurations
by the training.  
However,  there is a maximal shear strain beyond which the system cannot be stabilized anymore.
Our results for the compression training seem to be consistent with
this argument.
We remark that if we apply the compression-decompression cycle more than once, $\alpha$ also increases with $\varphi_\mathrm{J}$.
In this case, $\varphi_\mathrm{J}$ increases from 
$0.8469$ to $0.8481$
 and $\alpha$ increases from $\alpha = 0.72 \pm 0.02$ to $\alpha = 0.79 \pm 0.02$
with $\varphi_{\MAX}=1.0$ and the number of the training cycle
is 50.

Another subtle but possibly important change by the mechanical training 
observed in Figure \ref{Fig_Sq_Compressed.png} is 
that the window in which hyperuniformity is observed
widens slightly as $\varphi_{\MAX}$ increases.
In other words, the cut-off wavevector, or the inverse of the
hyperuniform length, $q_{\HU}$, below which $\chi(q)$ bends up, becomes
smaller.
This may be the consequences of the stabilization of the jamming
configurations by mechanical training, which is consistent with the argument in Ref.~\cite{Godfrey2018prl} that the 
ideal hyperuniformity with
$\chi(q)\rightarrow 0$ as $q\rightarrow 0$ is realized only in the
ideally stable configurations. 
However, the changes are too small to be conclusive. We leave more
systematic analysis using the larger systems and with different
training/annealing protocols for future work.

\section{Hyperuniformity of oscillatory sheared systems below the jamming transition point}
\label{sec:5}

In this section, we turn our attention to the system below the jamming
transition point and study the geometrical properties of the configurations under
mechanical training. 
\begin{figure}[htb]
\includegraphics[width=1.0\columnwidth,angle=-0]{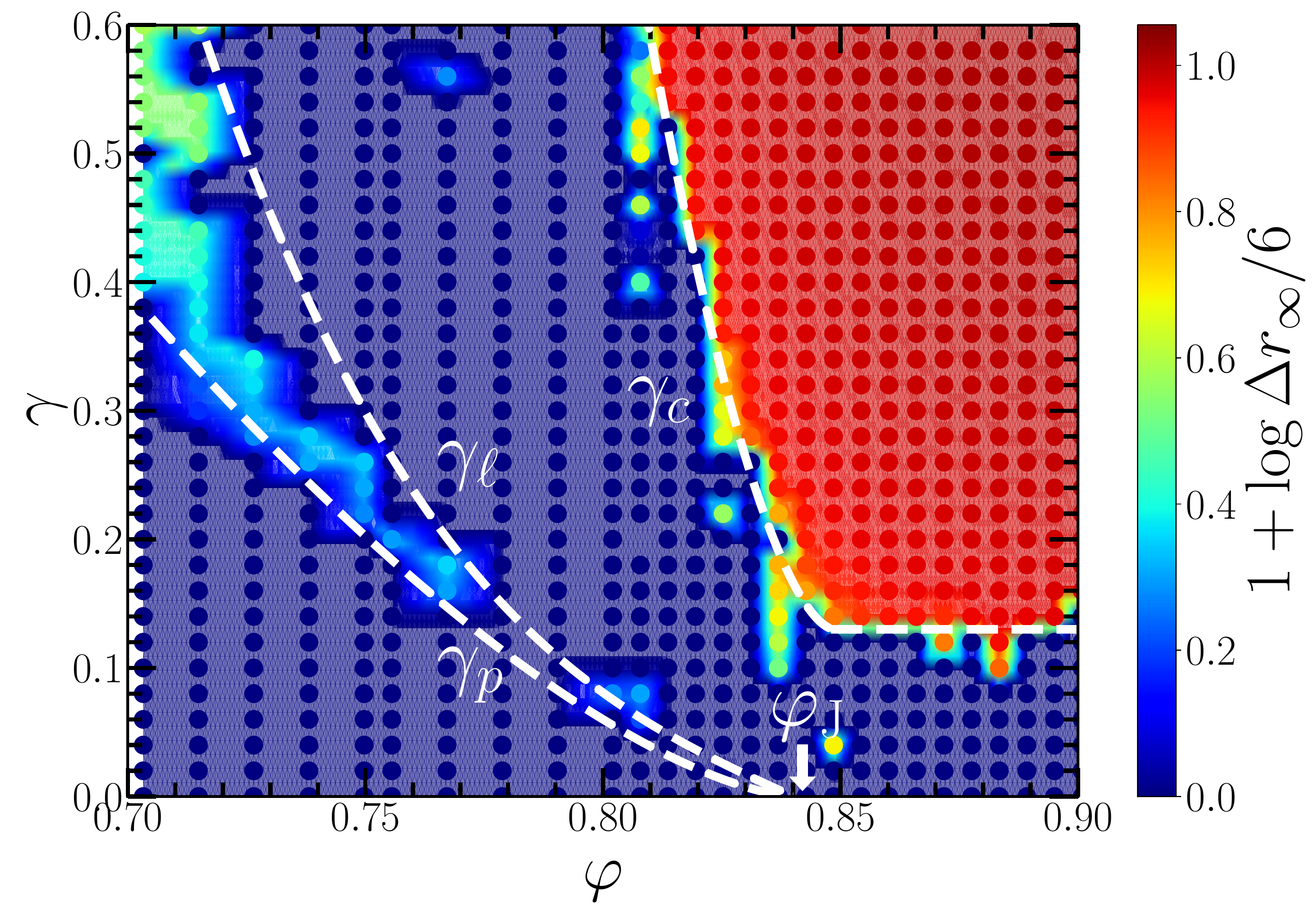}
\caption{The nonequilibrium phase diagram of the reversible-irreversible phase transition in the vicinity of the jamming point. 
The heat map of the average particle displacement per oscillation cycle in the stationary state,
 $\Delta r_{\infty}$ is plotted in the $(\varphi, \gamma)$-plane~\cite{Nagasawa2019sm}. 
The $\gamma_p$ and $\gamma_\ell$ lines separate the point-reversible
 phase, quasi-irreversible phase, and the loop-reversible phase, respectively.
$\gamma_c$ represents the reversible-irreversible transition line. 
The white-broken lines are guides for the eyes. 
 }
\label{Fig_RIRphase.jpg}
\end{figure}
Instead of compression, we apply the quasi-static oscillatory shear to the system.

The athermal and  periodically driven systems attract much attention
recently in the context of the nonequilibrium phase
transition in
the dilute regime far below the jamming transition point~\cite{Corte2008,Hexner2015prl,Tjhung2016jsmte,Zheng2021prl}.   
When periodically sheared, the athermal particles
originally distributed randomly move with the shear flow, collide with
other particles, and then redistribute after collisions. 
If the amplitude of the shear cycle, $\gamma$, is small, the particles find the
optimal configurations after multiple oscillatory shear cycles, 
in which all particles return to the same positions after every
cycle. 
If $\gamma$ exceeds a threshold value $\gamma_c$, the particles never come
back to the original positions and start diffusing over the space. 
It is called the reversible-irreversible (RI) transition and this transition is now
believed to belong to universality class either of the directed
percolation or conserved directed percolation (Manna) classes~\cite{Lubeck2004ijmpb}. 
Furthermore, the density fluctuations are suppressed and the system
becomes hyperuniform near the transition point.
The spectrum exhibits a power-law behavior $\chi(q)\sim q^{\alpha}$ at
small wavevectors with a universal exponent 
$\alpha \approx 0.45$ for $d=2$~\cite{Hexner2015prl,Tjhung2016jsmte,Zheng2021prl}  
\footnote{Tjhung {\it et al.} argued that $\alpha$ crosses over
to 1 at very small $q$~\cite{Tjhung2016jsmte}. } and $\alpha \approx
0.25$ for $d=3$\cite{wilken2020prl}.
The hyperuniformity of the RI transition is also studied in a
different experimental setup~\cite{wang2018nat}.

A natural question is how the nature of the RI transition at low-density
regime changes if the density is below but close to the jamming
transition point. 
Recently, the RI transition in the vicinity of $\varphi_\mathrm{J}$ has been
studied under the quasi-static oscillatory shear cycles in both 
$d=2$~\cite{Schreck2013pre,Nagasawa2019sm} and
$d=3$~\cite{Vinutha2016natp,Vinutha2016jsmte} slightly below 
and slightly above $\varphi_\mathrm{J}$~\cite{Das2020pnas}. 
Above $\varphi_\mathrm{J}$, the RI transition point $\gamma_c$ is almost identical
with the yielding transition point $\gamma_{Y}$~\cite{Kawasaki2016pre}, which is natural as the microscopic
trajectories of particles would be reversible in the elastic region 
and the trajectories become diffusive beyond
the yielding transition point where the system is fluidized. 
The exception is the critical region close to $\varphi_\mathrm{J}$ where the elastic response becomes nonlinear~\cite{Otsuki2014pre, Dagois-Bohy2017sm}. 

The nature of the RI transition slightly below $\varphi_\mathrm{J}$
is far richer. 
The $\gamma_c$-line suddenly increases as $\varphi$ departs below
$\varphi_\mathrm{J}$ (see Figure~\ref{Fig_RIRphase.jpg}). 
$\gamma_c$, however, is not vertical against $\varphi$ at $\varphi_\mathrm{J}$. 
The irreversible phase seeps into $\varphi < \varphi_\mathrm{J}$ and the
large $\gamma$ region. 
Interestingly, the microscopic and short-range structures of particles
along the $\gamma_c$-lines below $\varphi_\mathrm{J}$ are 
very similar to those of the jammed
configurations of the frictional particles~\cite{Vinutha2016natp}.
The reversible phase well below $\gamma_c$ is also unique in this
density regime~\cite{Schreck2013pre,Nagasawa2019sm}. 
If $\gamma$ is small enough, the trajectory of particles per cycle in
the stationary state is characterized by a straight line formed by a
shear convection, as the particles experience no collision with others
on the way forward and backward during one shear-cycle. 
We refer to this phase as the point-reversible phase since the
non-affine trajectories of the particles are just points. 
As $\gamma$ increases beyond a threshold value, $\gamma_{p}$, the
trajectories do not converge to a stationary state within the simulation
time windows. 
At larger $\gamma=\gamma_\ell$, the trajectories become reversible
again, but the nature of the trajectories change qualitatively. 
The particles experience multiple collisions and those particles
experience jaggy trajectories before traveling back to the original positions per shear-cycle. 
This phase exists until $\gamma$ crosses $\gamma_c$, beyond which the
trajectories become irreversible.
We call the phase between $\gamma_{\ell} \leq \gamma \leq \gamma_c$ the
loop-reversible phase. 
$\gamma_\ell$ is well below $\gamma_c$ and it converges to 0 as $\varphi$
approaches $\varphi_\mathrm{J}$.

The relaxation ``time'', $\tau_{c}$, which is defined by the number of oscillatory cycles
required for the system to reach the stationary state increases sharply
as $\gamma$ approaches $\gamma_{p}$~\cite{Schreck2013pre,Nagasawa2019sm}
and the order parameter such as the fraction of the irreversible particles
or the average particle displacement per cycle $\Delta r_{\infty}$ in 
the stationary state seems to increase continuously, in the stark contrast
with the RI transition at $\gamma_c$, where $\Delta r_{\infty}$ jump from
zero to a finite value discontinuously. 
We reported that there is a narrow strip of the irreversible phase just
above $\gamma_p$ in the previous study~\cite{Nagasawa2019sm}, where the
particles never reach the reversible state and thus 
$\Delta r_{\infty}$ remains small but finite (see the narrow region
between the $\gamma_p$ and $\gamma_\ell$ lines in Figure~\ref{Fig_RIRphase.jpg}). 
It may be a natural consequence of the diverging relaxation time
which exceeds the simulation time windows. But close inspections of the
loop trajectories indicate that the particles keep migrating over many
cycles with little sign to relax to a stationary state.  
The nature of this peninsula is still elusive and we shall name the region as the quasi-irreversible phase. 
Due to these complexities, the identification of the precise position of
the transition is challenging, which is the reason why the $\gamma_p$ and
$\gamma_\ell$ lines in Figure~\ref{Fig_RIRphase.jpg} are not sharply defined.

We investigate the density fluctuations in the vicinity of this
point-to-loop transition point $\gamma_p$ for the two-dimensional binary mixture.
In the simulation, we fix the density at $\varphi=0.82$, slightly below
$\varphi_\mathrm{J}\approx 0.842$. 
Note that, at this density, the irreversible phase was not clearly
observed and $\gamma_p \approx \gamma_\ell$. 
The box size is $L= 42$ and the number of samples is from 300 to 900, depends on $\gamma$.
We place the system under quasi-static oscillatory shear of various
amplitudes $\gamma$. 
In the quasi-static shear protocol, 
the system is deformed by the small shear $|\Delta \gamma|=10^{-3}$
incrementally. 
At each incremental step, the system's configuration
is relaxed by FIRE algorithm~\cite{Bitzek2006prl}.  
Note that in the FIRE algorithm, the finite mass is endowed to every particle 
and the trajectories towards the energy minima always overrun slightly
due to the inertia, which inevitably enhances the
chance for the particles to collide with others during the minimization processes. 
Thus, the particles' position after one shear cycle shifts slightly from
the original position.
We still use the FIRE algorithm instead of the conventional steepest descent method because it is far faster and computationally cheaper than the latter. 
We carefully checked the configurations generated by both algorithms and found that the difference is negligible as long as the density
fluctuations are concerned.
\begin{figure}[htb]
\includegraphics[width=1.0\columnwidth,angle=-0]{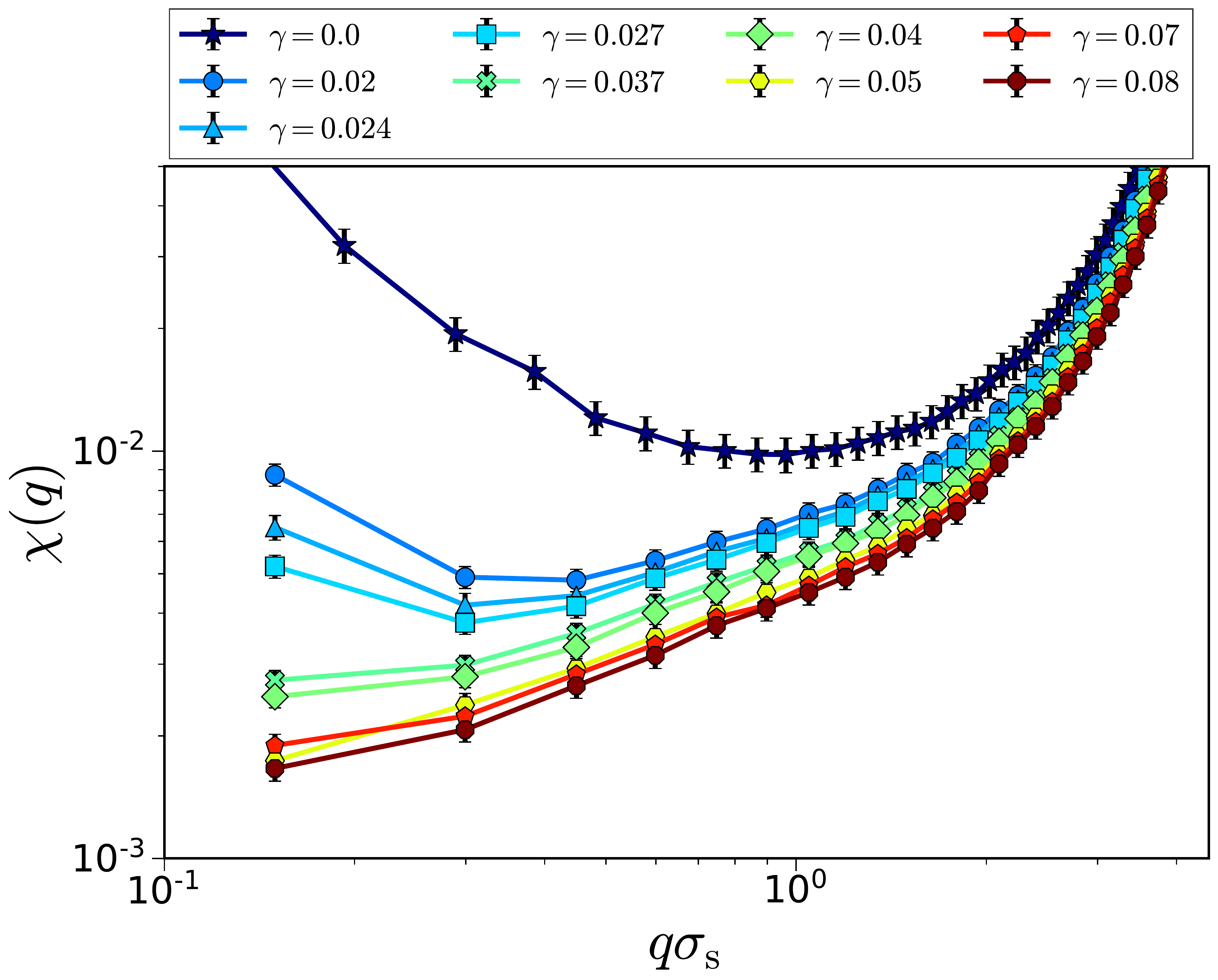}
\caption{$\chi(q)$ for various shear amplitude $\gamma$ at $\varphi=0.82$. 
At this density, the transition from the point to loop reversible phase takes place around $\gamma_p\approx 0.04$.
}
\label{Fig_sheared_Chiq.jpg}
\end{figure}

Figure~\ref{Fig_sheared_Chiq.jpg} shows $\chi(q)$ defined by
Eq.~\eqref{eq:chiq} for $\gamma = 0 \sim 0.08$. 
Note that $\gamma_p \approx 0.04$ at this density.
When $\gamma=0$, no sign of hyperuniformity is observed. $\chi(q)$
goes up at small $q$, reflecting that the particles' configuration is
random and it tends to obey the Poissonian statistics for which 
$\chi(q\rightarrow 0) =1$. 
However, as $\gamma$ increases, the upward bents are gradually suppressed.
Eventually, the line converges to a single curve which is algebraic.
This convergence takes place around $\gamma \approx 0.05$, which is very
close to $\gamma_p$ .
Surprisingly, the exponent of the curve $\chi(q) \sim q^{\alpha}$ is
$\alpha \approx 0.642\pm 0.014$ at $\gamma=0.05$. 
This exponent is close to that of the poorly trained system discussed in Section~\ref{sec:3}.

\begin{figure}[htb]
\includegraphics[width=1.0\columnwidth,angle=-0]{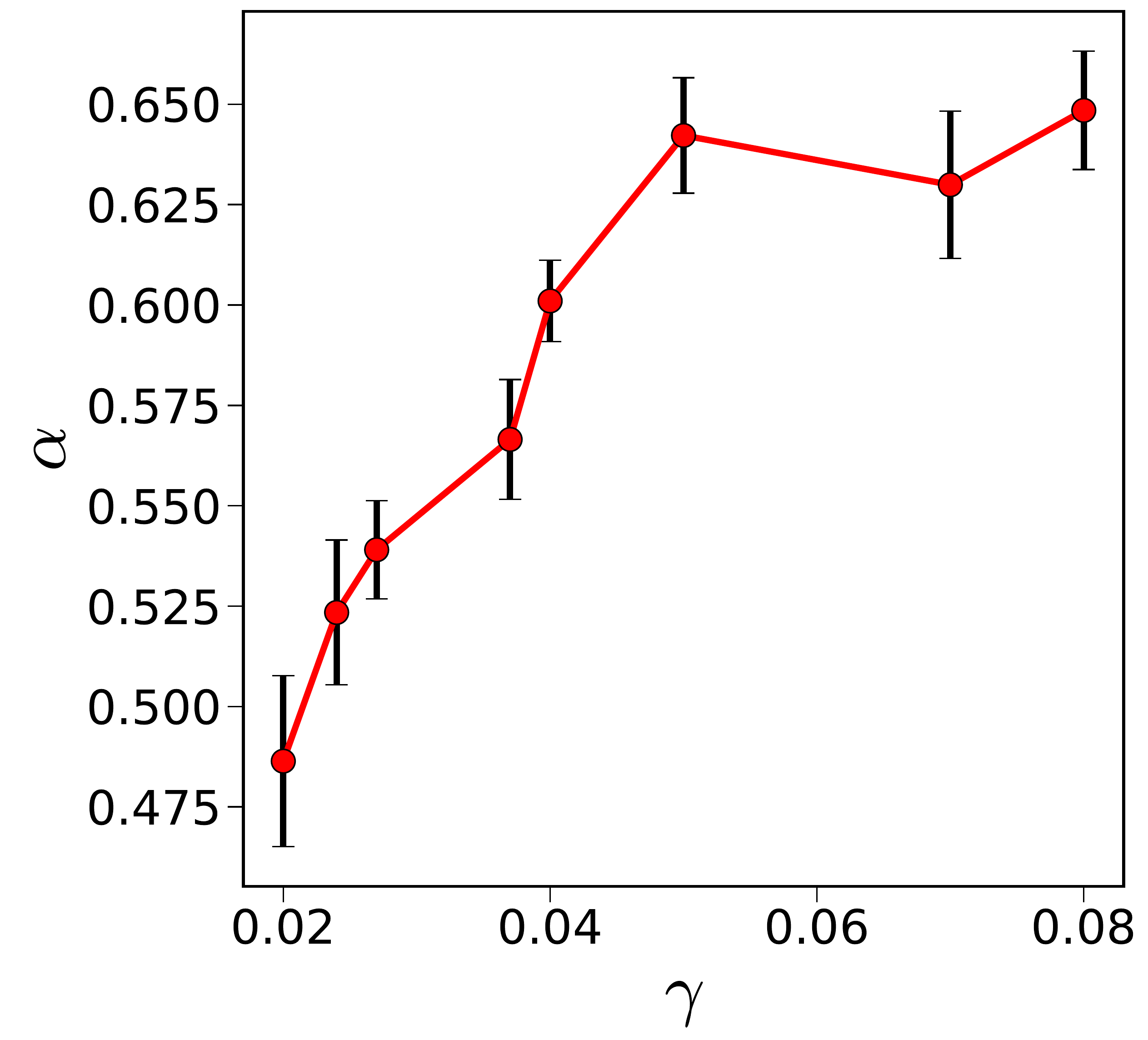}
\caption{$\gamma$-dependence of $\alpha$.
$\alpha$ monotonically increases with $\gamma$ up to 
$\gamma =0.05$. This value is close to $\gamma_p \approx 0.04$ at $\varphi= 0.82$.}
\label{Fig_sheared_alpha.jpg}
\end{figure}

We plot the $\gamma$-dependence of $\alpha$ in Figure~\ref{Fig_sheared_alpha.jpg}. 
$\alpha$'s are extracted by fitting by power-laws for the wavevector windows whose
size varies depending on $\gamma$'s. 
At small $\gamma$ regime, the fitting is less trustworthy because 
the power-law windows are narrow, if they exists.
Surprisingly, $\alpha$ becomes constant at large $\gamma$
and hyperuniformity persists well above the point-to-loop transition point, $\gamma_p$.
This result is in stark contrast with
the RI phase transition in the dilute regime where hyperuniformity
is observed only in the close vicinity of the transition point, $\gamma_c$. 
We emphasize that hyperuniformity observed here is distinct from that
observed for a similar density window but at a much larger shear amplitude
near $\gamma_c$ ($\gg \gamma_p$), 
where the discontinuous RI transition is observed~\cite{Vinutha2016jsmte}. 
Their exponent for the system under the strong oscillatory shear in
$d=3$ was $\alpha \approx 0.45$.
The result in Figure~\ref{Fig_sheared_Chiq.jpg}, on the other hand,
demonstrates that the sheared system with very small perturbation
inherits the geometrical properties of the configuration exactly at the
jamming point.
Contrary to the system trained by compression at 
$\varphi > \varphi_\mathrm{J}$, the exponent $\alpha$ remains constant for
large amplitude beyond $\gamma_{p}$.
Here we studied only one density $\varphi=0.82$. 
It is important to check whether this hyperuniform
behavior near the jamming transition crosses over to that near the DP
universality class at very low densities, where $\alpha \approx 0.45$.


\section{Summary}
\label{sec:summary}

In this paper, we shed light on the several facets of the geometrical
properties of the jammed or nearly jammed configurations of a two-dimensional binary mixture of particles with harmonic potential under nonequilibrium perturbation.
We primarily focused on hyperuniformity of the system by
monitoring the wavevector dependent compressibility $\chi(q)$ defined by
Eq.~\eqref{eq:chiq}. 
We first revisited hyperuniformity of unperturbed jammed
configuration.  The observation of $\chi(q)\sim q^{\alpha}$ with the
exponent of $\alpha \approx 0.63$ for $d=2$ instead of 
$\alpha \approx 1$ for the three-dimensional counterparts seems to be
unnoticed to the best of our knowledge, though the data reported in the
literature is consistent with those shown in Figure~\ref{fig:1}.
We then placed the jammed configurations under a compression cycle
from $\varphi < \varphi_\mathrm{J0}$ to $\varphi =\varphi_{\MAX}$. 
$\varphi_\mathrm{J}$ after this training cycle is slightly larger than
$\varphi_\mathrm{J0}$ of the poorly trained system.  
We found that the number of rattlers increased as $\varphi_\mathrm{J}$ increase.
This result indicates that the increases of $\varphi_\mathrm{J}$ do not
lead to the increase of the packing fraction of particles that sustain
the force network of the jammed structures.
We also find that the larger particles are responsible for the increase
in the number of rattlers. 
It is puzzling because our intuition tells us that 
the large particles would be easier to form 
the contact networks in a more stable configuration than the smaller ones
which would tend to find themselves in the free volume pockets.
More analysis of the system size dependence and the detailed microscopic
studies on the spatial correlations of the rattlers are necessary to
resolve this conundrum. 
Another finding is that the exponent of hyperuniformity $\alpha$ increases with $\varphi_{\MAX}$
from $\alpha \approx 0.63$ up to as large as $\alpha \approx 0.74$.
The value does not increase beyond this value with larger $\varphi_{\MAX}$.
Since mechanical training plays similar roles as thermal
annealing, our results imply that the suppression of the density
fluctuations should be observed for the jammed configurations generated
from the thermally annealed fluids. 
Furthermore, the wavevector $q_{\HU}$, the inverse of the
hyperuniform length, below which $\chi(q)$ ceases to
decrease and bends upward also tends to shift to lower values. The
results shown in Figure~\ref{Fig_Sq_Compressed.png} are not conclusive but this may suggest
that hyperuniformity extends to the large length scales as the system
is more trained and finds more stable configurations. 
This is consistent with the scenario proposed by Godfrey {\it et al.}~\cite{Godfrey2018prl}.

We also expect to observe qualitatively similar results when the system
is cyclically sheared instead of the compression cycles above $\varphi_\mathrm{J}$. 
Recently, Das {\it et al.}~\cite{Das2020pnas} have shown that
$\varphi_\mathrm{J}$ increases under the quasi-static oscillatory shear when
the shear strain amplitude is small, much the same way that $\varphi_\mathrm{J}$
increases with the compression cycle. 
It would be interesting to study hyperuniformity along
$\varphi_\mathrm{J}$-line, which increases with the shear amplitude $\gamma$ and
see if the exponent $\alpha$ increases and $q_{\HU}$ decreases with the
shear training~\cite{Shukawa-unpub}.  
In Section 5, we have carried out the quasi-static oscillatory shear
experiment at $\varphi =0.82$ that is below $\varphi_\mathrm{J}=0.842$. 
The system is not hyperuniform when $\gamma=0$, but as the shear
amplitude increases, the large-scale density fluctuations becomes suppressed
gradually and eventually become hyperuniform with the same
exponent $\alpha \approx 0.64$ as the poorly trained configuration at 
the jamming transition point $\varphi_\mathrm{J}$. 
The convergence to hyperuniform state takes place when 
$\gamma$ crosses the transition point $\gamma_p \approx \gamma_\ell$.
Hyperuniformity observed here is distinct from that observed at the
reversible-to-irreversible transition point $\gamma_c$ at larger 
$\gamma$~\cite{Vinutha2016natp,Vinutha2016jsmte}.
Figure~\ref{Fig_RIRphase.jpg} hints that the
$\gamma_c$-line near the jamming transition point
does not extrapolate and connect smoothly to the RI transition line at the low
density limit, where the {\it bona fide} nonequilibrium transition 
of the DP (or Manna) universality classes are expected. 
Connecting the RI transition lines at the two disparate density
regions are important future work.

In this paper, we displayed the results only for a few representative state
points with moderate system sizes. 
All mechanical perturbation used here is quasi-static protocols and 
requires heavy computational loads as well as many samplings to obtain
the good statistics.
Exploring systematically more state points with larger system sizes 
will be necessary to reveal the whole pictures of 
the rich and anomalous geometrical properties of the nonequilibrium
configuration near the jamming transition point.

\begin{acknowledgements}
Discussion with Srikanth Sastry and  Misaki Ozawa is very fruitful. 
This work was financially supported by KAKENHI
Grants 18H01188, 19H01812, 19K03767, 20H05157, and
20H00128.
\end{acknowledgements}

\normalem 
\bibliographystyle{spphys_modify}     

\end{document}